\documentclass{article} 
\usepackage{graphicx}  
\usepackage{amsmath}   
\usepackage[compress]{cite}
\usepackage{amssymb}   
\usepackage{bm} 
\usepackage{dcolumn}
\usepackage{color}
\usepackage{mathrsfs}
\usepackage{amsfonts}
\usepackage{varioref}
\RequirePackage[colorlinks,citecolor=blue,urlcolor=magenta,linkcolor=blue]{hyperref}
\labelformat{section}{Section #1} 
\labelformat{subsection}{Section #1} 
\labelformat{subsubsection}{Section #1}
\labelformat{subsubsubsection}{Section #1}
\labelformat{equation}{Eq.~(#1)} 
\labelformat{figure}{Fig.~#1} 
\labelformat{subfigure}{Fig.~\thefigure#1} 
\labelformat{table}{Tab.~#1} 
\labelformat{appendix}{Appendix #1}
\title{\bf Bound on photon circular orbits in general relativity and beyond}
\author{Sumanta Chakraborty
\footnote{sumantac.physics@gmail.com}~\footnote{tpsc@iacs.res.in}\\
{\small{School of Physical Sciences, Indian Association for the Cultivation of Science, Kolkata-700032, India}}}
\begin{document}
  
\maketitle
\begin{abstract}
Existence of a photon circular orbit can tell us a lot about the nature of the underlying spacetime, since it plays a pivotal role in the understanding of the characteristic signatures of compact objects, namely the quasi-normal modes and shadow radius. For this purpose, determination of the location of the photon circular orbit is of utmost importance. In this work, we derive bounds on the location of the photon circular orbit around compact objects within the purview of general relativity and beyond. As we have explicitly demonstrated, contrary to the earlier results in the context of general relativity, the bound on the location of the photon circular orbit is not necessarily an upper bound, rather depending on the matter content it is possible to arrive at a lower bound as well. This have interesting implications for the quasi-normal modes and shadow radius, the two key observables related to the strong field tests of gravity. Besides discussing the bound for higher dimensional general relativity, we have also considered how the bound on the photon circular orbits gets modified in the braneworld scenario, for pure Lovelock and general Lovelock theories of gravity. Implications of these results for compact objects have also been discussed.
\end{abstract}
\newpage
\tableofcontents
\section{Introduction}

Detection of gravitational waves from the merger of binary black holes and observing shadow of a supermassive black hole have opened up unprecedented avenues to test gravitational theories in the strong field regime \cite{Abbott:2016blz,TheLIGOScientific:2016wfe,TheLIGOScientific:2016src,Abbott:2016nmj,LIGOScientific:2019fpa,Abbott:2020jks,Psaltis:2020lvx,Akiyama:2019cqa}. This is pivotal in the search for theories beyond general relativity, though the most successful theory so far in describing the gravitational interaction in very many different length scales, there are regimes where even general relativity fails \cite{PhysRevLett.14.57,PhysRevD.14.2460,Christodoulou:1991yfa,Will:2005va,Berti:2015itd}. Such a regime corresponds to the region near the singularity, both for black holes and cosmological spacetimes. Additionally --- (i) presence of late time cosmic acceleration \cite{RevModPhys.61.1,Carroll:2000fy,Padmanabhan:2002ji,Elizalde:2004mq,Cognola:2006eg}, (ii) violation of strong cosmic censorship conjecture \cite{Cardoso:2017soq,Dias:2018etb,PhysRevD.97.104060,Rahman:2018oso,Rahman:2020guv,Mishra:2020jlw}, (iii) flat rotation curves of galaxies \cite{Cowie:1986eb,Binney:1987,Persic:1995ru,Borriello:2000rv}, among others, cannot be explained by classical general relativity with traditional matter content. These require either postulating existence of some exotic matter fields or theories beyond general relativity \cite{Copeland:2006wr,Ade:2013zuv,Nojiri:2010wj,DiValentino:2021izs}, or, some quantum effects must be taken into account \cite{Hollands:2019whz}. All of which motivates one to look for alternatives of general relativity, which may overcome these issues and can possibly provide a resolution to the singularity problem. In the weak field regimes, most of the predictions of these alternative theories are consistent with that of general relativity and often provide weak experimental/observational bounds \cite{Chakraborty:2012sd,Bhattacharya:2016naa,Iorio:2012cm,Capozziello:2006jj}. However, predictions of these alternative theories will start to differ significantly from that of general relativity in the strong field regime, i.e., in the region around the black hole horizon and the photon sphere. Thus it is of utmost importance to search for these theories from the gravitational waves and shadow measurements, probing the near horizon regime of black holes.   

Intriguingly, the photon sphere plays a central role in determining the characteristic properties of black holes. The shadow diameter, for example, is associated with the capture cross-section of a black hole and is related to the energy and angular momentum of a photon moving along the photon circular orbit \cite{Abdujabbarov:2015rqa,Chakraborty:2016lxo,Abdujabbarov:2017pfw,Mishra:2019trb,Abdikamalov:2019ztb,Banerjee:2019nnj,Toshmatov:2021fgm,Jusufi:2019caq,Li:2021mnx}. Similarly, the quasi-normal modes, which are unique characteristics of a black hole, also depends on the physical properties of the photon sphere in the eikonal limit \cite{Cardoso:2008bp}. Therefore, all the strong field tests of gravity available to us at the present instance are intimately tied up with the properties, in particular, the location of the photon sphere. Another important aspect is the universality of the photon sphere, i.e., the gravitational wave and shadow observations cannot provide conclusive evidence regarding the central object being a black hole, but they do agree on the existence of a photon sphere. Thus the central compact object, which may, or, may not be a black hole, must have a photon sphere \cite{Cardoso:2019rvt,Maggio:2017ivp,Dey:2020lhq,Maggio:2020jml,Raposo:2018rjn}. Such compact, but non-black hole objects, known as exotic compact objects (or, ECOs), require exotic matter fields for their support and have distinct signatures in the quasi-normal mode spectrum. However, they all have the photon sphere as a common entity \cite{Chakraborty:2021gdf,Dey:2020pth,Hod:2020pim,Hod:2018kql,Ghosh:2021txu}. Such versatility of the existence of photon sphere motivates us to study it in more detail in general relativity and theories beyond the same. 

Even though looking for theories beyond general relativity is necessary to understand the nature of gravity in the strong field regime, it is difficult to break the degeneracy among those theories, since there are infinite number of them. The situation is improved by invoking the Ostr\"{o}gradsky instability \cite{Woodard:2015zca}, thereby eliminating all the higher curvature gravitational theories yielding more than second order gravitational field equations. These leave a handful of choices --- (i) higher dimensional/braneworld models of gravity \cite{Overduin:1998pn,Csaki:2004ay,PerezLorenzana:2005iv,Chamblin:1999by,Emparan:1999wa,Shiromizu:1999wj}, (ii) Lovelock theories of gravity \cite{Chakraborty:2015wma,Parattu:2015gga}, (iii) scalar-tensor theories of gravity, e.g., $f(R)$ theories, Horndeski theories etc. \cite{Nojiri:2010wj,Clifton:2011jh,Babichev:2016rlq,Sotiriou:2008rp}. The modified theories, listed above, provide unique predictions for the quasi-normal modes of black holes and of course, for the photon sphere. In the present work we wish to explore the implications of these beyond general relativity theories on the location of the photon sphere, which in turn will affect both the quasi-normal modes and the shadow radius.  

The paper is organized as follows: In \ref{Ph_GR_CC_HD} we present the bound on the photon circular orbit in the context of higher dimensional general relativity, which has been contrasted to the corresponding situation in the braneworld scenario, discussed in \ref{Ph_Brane_CC}. Subsequently, we have derived the corresponding bound on the photon circular orbit for pure Lovelock theories in \ref{Ph_pureLovelock}. Generalization of the same to general Lovelock theories has been presented in \ref{Ph_genLovelock}, with the corresponding scenario for the Einstein-Gauss-Bonnet theory discussed in \ref{Ph_EGB} as a warm-up exercise. Applications of these results for black hole quasi-normal modes and shadow have been presented in \ref{Ph_App}, before concluding with certain remarks and future directions. 

\emph{Notations and Conventions:} In the present work we set all the fundamental constants $c$, $G$ and $\hbar$ to unity and shall follow the mostly positive signature convention, such that the flat spacetime metric takes the following form, $\eta_{\mu \nu}=\textrm{diag.}(-1,+1,+1,\cdots)$. Additionally, Greek indices $\alpha,\beta,\mu,\cdots$ run over the four dimensional coordinates and Roman indices $a,b,c,\cdots$ run over all the spacetime coordinates. 
\section{Bound on photon circular orbits in general relativity}\label{Ph_GR_CC_HD}

It is always wise to first understand the basic premise of a problem, which motivates us to study the bound on the photon circular orbit in general relativity itself, in this section. This can be considered as a warm up to the subsequent discussions involving theories beyond general relativity. In what follows, we will derive the relevant bound on the photon circular orbit, for generic static and spherically symmetric spacetimes in general relativity, with arbitrary spacetime dimensions. The result can then be easily specialized to the case of four spacetime dimensions. As a starting point, we will assume the following metric ansatz for describing a static and spherically symmetric $d$-dimensional spacetime in general relativity, which reads,
\begin{align}\label{Eq_Ph_01}
ds^{2}=-e^{\nu(r)}dt^{2}+e^{\lambda(r)}dr^{2}+r^{2}d\Omega_{d-2}^{2}~.
\end{align}
Substitution of this metric ansatz in the Einstein's field equations, with anisotropic perfect fluid as the matter source, yields the following field equations for the unknown functions, $\nu(r)$ and $\lambda(r)$, in $d$ spacetime dimensions, 
\begin{align}
r\lambda' e^{-\lambda}+\left(d-3\right)\left(1-e^{-\lambda}\right)&=\left(8\pi \rho +\Lambda \right)r^{2}~,
\label{Eq_Ph_02a}
\\
r\nu' e^{-\lambda}-\left(d-3\right)\left(1-e^{-\lambda}\right)&=\left(8\pi p-\Lambda \right)r^{2}~,
\label{Eq_Ph_02b}
\end{align}
where, `prime' denotes derivative with respect to the radial coordinate $r$. It must be noted that we have incorporated the cosmological constant $\Lambda$ in the above analysis. The differential equation for $\lambda(r)$, presented in \ref{Eq_Ph_02a}, can be immediately integrated, since the left hand side of the equation is expressible as a total derivative term, except for some overall factor, leading to,
\begin{align}\label{Eq_Ph_04}
e^{-\lambda}=1-\frac{2m(r)}{r^{d-3}}-\frac{\Lambda}{(d-1)}r^{2}~;
\qquad 
m(r)=M_{\rm H}+4\pi \int _{r_{\rm H}}^{r}dr' \rho(r')r'^{d-2}~.
\end{align}
Here, $M_{\rm H}$ denotes the mass of the black hole, with its horizon radius being $r_{\rm H}$. This situation is very much similar to the case of black hole accretion, where $\rho(r)$ and $p(r)$ are respectively the energy density and pressure of matter fields accreting onto the black hole spacetime. Being spherically symmetric, we can simply concentrate on the equatorial plane and the photon circular orbit on the equatorial plane arises as a solution to the algebraic equation, $r\nu'=2$. Analytical expression for $\nu'$ can be derived from \ref{Eq_Ph_02b}, whose substitution into the equation $r\nu'=2$, yields the following algebraic equation,
\begin{align}\label{Eq_Ph_06}
\left(8\pi pr^{2}-\Lambda r^{2}\right)+(d-3)\left(1-e^{-\lambda}\right)=2e^{-\lambda}~,
\end{align}
which is independent of $\nu(r)$ and dependent only on $\lambda(r)$ and matter variables. At this stage it will be useful to define the following quantity,
\begin{align}\label{Eq_Ph_07}
\mathcal{N}_{\rm gr}(r)\equiv -8\pi pr^{2}+\Lambda r^{2}-(d-3)+(d-1)e^{-\lambda}~,
\end{align}
such that on the photon circular orbit $r_{\rm ph}$, we have $\mathcal{N}_{\rm gr}(r_{\rm ph})=0$, which follows from \ref{Eq_Ph_06}. Using the solution for $e^{-\lambda}$, in terms of the mass $m(r)$ and the cosmological constant $\Lambda$, from \ref{Eq_Ph_04}, the function $\mathcal{N}_{\rm gr}(r)$, defined in \ref{Eq_Ph_07}, yields,
\begin{align}\label{Eq_Ph_12}
\mathcal{N}_{\rm gr}(r)=-8\pi pr^{2}+\Lambda r^{2}-(d-3)+(d-1)\left(1-\frac{2m(r)}{r^{d-3}}-\frac{\Lambda}{(d-1)}r^{2}\right)
=2-2(d-1)\frac{m(r)}{r^{d-3}}-8\pi pr^{2}~,
\end{align}
which is independent of the cosmological constant $\Lambda$. It is further assumed that both the energy density $\rho(r)$ and the pressure $p(r)$ decays sufficiently fast, so that, $pr^{2}\rightarrow 0$ and $m(r)\rightarrow \textrm{constant}$ as $r\rightarrow \infty$. Thus from \ref{Eq_Ph_12} it immediately follows that,
\begin{align}\label{Eq_Ph_13}
\mathcal{N}_{\rm gr}(r\rightarrow \infty)=2~.
\end{align}
Note that this asymptotic limit of $\mathcal{N}_{\rm gr}(r)$ is independent of the presence of higher dimension, as well as of the cosmological constant and will play an important role in the subsequent analysis. 

It is possible to derive a few interesting relations and inequalities for the matter variables and also for the metric functions, on and near the horizon. The first of such relations can be derived by adding the two Einstein's equations, written down in \ref{Eq_Ph_02a} and \ref{Eq_Ph_02b}, which yields,
\begin{align}\label{Eq_Ph_08}
e^{-\lambda}\left(\frac{\nu'+\lambda'}{r}\right)=8\pi (p+\rho)~.
\end{align}
This relation must hold for all possible choices of the radial coordinate $r$, including the horizon. The horizon, by definition, satisfies the condition $e^{-\lambda(r_{\rm H})}=0$, thus if $\nu'+\lambda'$ is assumed to be finite at the location of the horizon, it follows that,
\begin{align}\label{Eq_Ph_09}
\rho(r_{\rm H})+p(r_{\rm H})=0~.
\end{align}
In addition to the above result, it also follows that $e^{-\lambda}\lambda'<0$ at the horizon location. This can be seen as follows: we have $e^{-\lambda}=0$ on the black hole horizon, while $e^{-\lambda}>0$ for $r_{\rm H}<r<r_{\rm C}$, where $r_{\rm H}$ is the black hole horizon and $r_{\rm C}$ is the cosmological horizon. Thus, $e^{-\lambda}$ is an increasing function at the black hole horizon and hence it follows that, $\partial _{r}e^{-\lambda}>0$ at $r=r_{\rm H}$, leading to $e^{-\lambda}\lambda '<0$ at $r=r_{\rm H}$. Thus from \ref{Eq_Ph_02a} we obtain the following inequality,
\begin{align}\label{Eq_Ph_10}
8\pi r_{\rm H}^{2}\rho(r_{\rm H})+\Lambda r_{\rm H}^{2}=r_{\rm H}\lambda'(r_{\rm H})e^{-\lambda(r_{\rm H})}+\left(d-3\right)<(d-3)~.
\end{align}
Using this inequality on the black hole horizon, along with the fact that $e^{-\lambda(r_{\rm H})}=0$, another inequality can be derived from \ref{Eq_Ph_07} involving $\mathcal{N}_{\rm gr}(r)$ on the black hole horizon, which reads,
\begin{align}\label{Eq_Ph_11}
\mathcal{N}_{\rm gr}(r_{H})=-8\pi p(r_{H})r_{H}^{2}+\Lambda r_{H}^{2}-(d-3)=8\pi \rho(r_{H})r_{H}^{2}+\Lambda r_{H}^{2}-(d-3)< 0
\end{align}
Therefore, being a monotonic function, it follows that $\mathcal{N}_{\rm gr}(r)$ is negative within the radial range $r_{\rm H}\leq r \leq r_{\rm ph}$. This result will be crucial in deriving the bound on the photon circular orbit subsequently. 

The next step is to write down the conservation relation of the energy momentum tensor. Note that we have expressed the temporal and radial components of the Einstein's equations in \ref{Eq_Ph_02a} and \ref{Eq_Ph_02b}, respectively, but we have not written down the angular components. The conservation of the energy momentum tensor will serve as a proxy for the same. In the context of higher dimensional spacetime, with anisotropic fluid, the conservation of the energy momentum tensor yields,
\begin{align}\label{Eq_Ph_14}
p'+\frac{\nu'}{2}\left(p+\rho \right)+\left(\frac{d-2}{r}\right)\left(p-p_{\rm T}\right)=0~,
\end{align}
where, $p_{\rm T}$ is the angular or, transverse pressure of the fluid, taken to be different from the radial pressure $p$. Substitution of the expression for $\nu'$ from the radial Einstein's equations, i.e., \ref{Eq_Ph_02b}, yields the following expression for $(dp/dr)$ in terms of $\mathcal{N}_{\rm gr}(r)$, 
\begin{align}\label{Eq_Ph_16}
p'(r)&=\frac{e^{\lambda}}{2r}\Big[\left(p+\rho\right)\mathcal{N}_{\rm gr}+2e^{-\lambda}\left\{-\rho+p+(d-2)p_{\rm T}\right\}-2de^{-\lambda}p\Big]~,
\end{align}
where \ref{Eq_Ph_07} has been used. Introducing the rescaled pressure $P(r)$, defined as, $P(r)\equiv r^{d}p(r)$, we obtain,
\begin{align}\label{Eq_Ph_17}
P'(r)=\frac{r^{d-1}e^{\lambda}}{2}\Big[\left(p+\rho\right)\mathcal{N}_{\rm gr}+2e^{-\lambda}\left\{-\rho+p+(d-2)p_{\rm T}\right\} \Big]~.
\end{align}
Assuming that, $\rho\geq 0$ everywhere, it is clear from \ref{Eq_Ph_09} that $p(r_{\rm H})$ and hence $P(r_{\rm H})\leq 0$. Further assuming that the trace of the energy momentum tensor, $-\rho+p+(d-2)p_{\rm T}$  is negative \cite{Hod:2020pim}, it follows that $P'(r_{\rm ph})<0$, since $\mathcal{N}_{\rm gr}(r_{\rm ph})=0$. Along identical lines and from \ref{Eq_Ph_11}, it further follows that $P'(r_{\rm H})<0$ as well. Thus one readily arrives at the following condition,
\begin{align}\label{Eq_Ph_18}
P'(r_{\rm H}\leq r \leq r_{\rm ph})<0~.
\end{align}
This suggests that $P(r)$ decreases as the radial distance increases from the horizon, located at $r_{\rm H}$ to the photon sphere, at $r_{\rm ph}$. This is because, $\mathcal{N}_{\rm gr}$ is negative in this radial distance range and so is the trace of the energy momentum tensor. Therefore, $p(r_{\rm H})\leq 0$, from which it follows that $p(r_{\rm ph})\leq 0$ as well. Thus from the result $\mathcal{N}_{\rm gr}(r_{\rm ph})=0$, substituted into \ref{Eq_Ph_12}, it follows that,
\begin{align}\label{Eq_Ph_19}
2-2(d-1)\frac{m(r_{\rm ph})}{r_{\rm ph}^{d-3}}\leq 0~.
\end{align}
Since $m(r_{\rm ph})<\mathcal{M}\equiv m(r\rightarrow \infty)$, where $\mathcal{M}$ is the ADM mass of the spacetime at infinity, we finally arrived at the desired bound on the radius $r_{\rm ph}$ of the photon circular orbit,
\begin{align}\label{Eq_Ph_20}
r_{\rm ph}\leq \left\{(d-1)\mathcal{M}\right\}^{1/(d-3)}~.
\end{align}
For $d=4$, the above inequality immediately suggests $r_{\rm ph}\leq 3\mathcal{M}$, which coincides with the result derived in \cite{Hod:2020pim}. It is worth mentioning that it is also possible, within the context of general relativity, to arrive at the above bound on the location of the photon circular orbit using the null energy conditions alone. Following \cite{Yang:2019zcn}, this requires one to define a new mass function,
\begin{align}
\mathbb{M}(r)&=\mathtt{m}(r)+\frac{4\pi}{(d-1)}r^{d-1}p-\frac{\Lambda r^{d-1}}{2(d-1)}=\frac{r^{d-3}}{2}\left(1-e^{-\lambda}\right)+\frac{4\pi}{(d-1)}r^{d-1}p-\frac{\Lambda r^{d-1}}{2(d-1)}
\nonumber
\\
&=\frac{r^{d-3}}{(d-1)}\left[1-\frac{1}{2}\left\{-(d-3)+(d-1)e^{-\lambda}-8\pi r^{2}p+\Lambda r^{2}\right\}\right]=\frac{r^{d-3}}{(d-1)}\left[1-\frac{\mathcal{N}_{\rm gr}}{2}\right]~,
\end{align}
where, $\mathtt{m}(r)$ is the Hawking-Geroch mass function, which for static and spherically symmetric $d$-dimensional spacetimes in general relativity takes the form, $(r^{d-3}/2)(1-e^{-\lambda})$. Note that on the photon sphere, $\mathbb{M}(r_{\rm ph})=r^{d-3}_{\rm ph}/(d-1)$ and on the horizon, $\mathbb{M}(r_{\rm H})>r^{d-3}_{\rm H}/(d-1)$. Moreover, derivative of this modified mass function $\mathbb{M}(r)$ takes the form,
\begin{align}
\mathbb{M}'(r)&=\frac{(d-3)r^{d-4}}{2}\left(1-e^{-\lambda}\right)+\frac{r^{d-3}\lambda'}{2}e^{-\lambda}+\frac{4\pi}{(d-1)}r^{d-1}p'+4\pi r^{d-2}p-\frac{\Lambda r^{d-2}}{2}
\nonumber
\\
&=\frac{(d-3)r^{d-4}}{2}\left(1-e^{-\lambda}\right)+\frac{\left(8\pi \rho+\Lambda\right)r^{d-2}}{2}-\frac{(d-3)r^{d-4}}{2}\left(1-e^{-\lambda}\right)+4\pi r^{d-2}p-\frac{\Lambda r^{d-2}}{2}
\nonumber
\\
&\hskip 1 cm +\frac{2\pi r^{d-2}e^{\lambda}}{(d-1)}\Big[\left(p+\rho\right)\mathcal{N}_{\rm gr}+2e^{-\lambda}\left\{-\rho+p+(d-2)p_{\rm T}\right\}-2de^{-\lambda}p\Big]
\nonumber
\\
&=\frac{2\pi r^{d-2}e^{\lambda}}{(d-1)}\left(p+\rho\right)\mathcal{N}_{\rm gr}+4\pi \rho r^{d-2}+4\pi r^{d-2}p+\frac{4\pi r^{d-2}}{(d-1)}\left\{-\rho+p+(d-2)p_{\rm T}\right\}-\frac{4\pi d p r^{d-2}}{(d-1)}
\nonumber
\\
&=\frac{2\pi r^{d-2}e^{\lambda}}{(d-1)}\left(p+\rho\right)\mathcal{N}_{\rm gr}+\frac{4\pi r^{2}(d-2)}{(d-1)}\left(\rho+p_{\rm T}\right)~,
\end{align}
where in the second line we have used \ref{Eq_Ph_16}. From our previous analysis we observe that $\mathcal{N}_{\rm gr}\geq 0$ for $r\geq r_{\rm ph}$ and the null energy condition ensures that $(\rho+p)$, as well as $(\rho+p_{\rm T})$ are positive definite quantities. Thus both the terms in the above expression for $\mathbb{M}'(r)$ is positive and hence it follows that $\mathbb{M}'(r)\geq 0$ as well. From the positivity of $\mathbb{M}'(r)$, for $r\geq r_{\rm ph}$, it follows that the mass function $\mathbb{M}(r)$ is an increasing function of the radial coordinate and hence $\mathbb{M}(r_{\rm ph})\leq \mathcal{M}$, which is the ADM mass of the spacetime. As mentioned earlier, on the photon sphere, we have $\mathbb{M}(r_{\rm ph})=r^{d-3}_{\rm ph}/(d-1)$ and hence \ref{Eq_Ph_20} follows immediately. Therefore, it is indeed possible to arrive at the desired bound using only the null energy condition. Note that the cosmological constant has no influence on the bound on the photon circular orbit, while the bound indeed depends on the spacetime dimensions. In what follows we will demonstrate that such bounds on the photon circular orbits indeed exists for theories of gravity beyond general relativity, though it is not necessarily a lower bound.  

\section{Bound on circular photon orbits in braneworld gravity}\label{Ph_Brane_CC}

In the previous section, we had derived an upper bound on the location of the photon circular orbit in the presence of a cosmological constant and higher spatial dimensions in the context of general relativity. In this section we will discuss a similar higher dimensional situation, but in the context of effective gravitational field theories, as observed by an observer on the four dimensional brane. As a consequence, the gravitational field equations on the brane will inherit additional corrections due to the presence of higher dimensions \cite{Chakraborty:2015taq,Chakraborty:2014xla}. In the subsequent discussion we will try to understand how these additional terms in the effective gravitational field equations may affect the bound on the photon circular orbit. 

To start with we assume a static and spherically symmetric metric ansatz as presented in \ref{Eq_Ph_01}, with $\nu(r)$ and $\lambda(r)$ as two unknown functions of the radial coordinate $r$ and the angular part $d\Omega _{d-2}^{2}$ is being replaced by $d\Omega _{2}^{2}$, as fit for the four dimensional braneworld scenario. In this context the temporal and the radial part of the field equations, with an effective four dimensional cosmological constant read,
\begin{align}
e^{-2\lambda(r)}\left(\frac{1}{r^{2}}-\frac{2\lambda'}{r}\right)-\frac{1}{r^{2}}&=-8\pi G_{4}\rho\left(1+\frac{\rho}{2\lambda _{\rm b}}\right)
-24\pi G_{4} \widetilde{U}(r)-\Lambda_{4}~,
\label{Eq_Ph_21a}
\\
e^{-2\lambda(r)}\left(\frac{2\nu'}{r}+\frac{1}{r^{2}}\right)-\frac{1}{r^{2}}&=8\pi G_{4}\left\{p+\frac{\rho}{2\lambda _{\rm b}}\left(\rho+2p\right) \right\}+8\pi G_{4} \left(\widetilde{U}+2\widetilde{P}\right)-\Lambda _{4}~.
\label{Eq_Ph_21b}
\end{align}
Here $\lambda _{\rm b}$ is the brane tension, $G_{4}$ is the four dimensional gravitational constant and $\Lambda _{4}$ is the brane cosmological constant. The other quantities, except the physical pressure $p$ and density $\rho$, correspond to the dark radiation $\widetilde{U}$ and dark pressure $\widetilde{P}$, respectively, inherited from the higher dimensional spacetime. These are derived from the projected bulk Weyl tensor $E_{\mu \nu}$ on the brane, such that, $\widetilde{U}=2G_{4}U(8\pi G_{4})^{-2}\lambda _{\rm b}^{-1}$, where $U=-(G_{4}/G_{5})^{2}E_{\mu \nu}u^{\mu}u^{\nu}$, with $G_{5}$ being the five dimensional gravitational constant and $u^{\mu}$ is the four velocity of a static observer in the spacetime. Similarly, the dark pressure term $\widetilde{P}$ can also be derived from the projected bulk Weyl tensor $E_{\mu \nu}$, such that, $\widetilde{P}=2G_{4}P(8\pi G_{4})^{-2}\lambda _{\rm b}^{-1}$, with $P$ being $(G_{4}/G_{5})^{2}E_{\mu \nu}r^{\mu}r^{\nu}$, where $r_{\mu}$ is orthogonal to the four-velocity of the static observer, such that, $r_{\mu}u^{\mu}=0$.  

Having discussed the content of the above equations in some detail, let us rewrite these gravitational field equations, i.e., \ref{Eq_Ph_21a} and \ref{Eq_Ph_21b}, such that we obtain the following ones,
\begin{align}
e^{-2\lambda(r)}\left(\frac{1}{r^{2}}-\frac{2\lambda'}{r}\right)-\frac{1}{r^{2}}&=-8\pi G_{4}\rho_{\rm eff}-\Lambda_{4};\qquad
\rho _{\rm eff}=\rho\left(1+\frac{\rho}{2\lambda _{\rm b}}\right)+3 \widetilde{U}(r)~,
\label{Eq_Ph_22a}
\\
e^{-2\lambda(r)}\left(\frac{2\nu'}{r}+\frac{1}{r^{2}}\right)-\frac{1}{r^{2}}&=8\pi G_{4}p_{\rm eff}-\Lambda _{4};\qquad
p_{\rm eff}=\left\{p+\frac{\rho}{2\lambda _{\rm b}}\left(\rho+2p\right) \right\}+\left(\widetilde{U}+2\widetilde{P}\right)~.
\label{Eq_Ph_22b}
\end{align}
As far as the transverse pressure is concerned, it is given by $p_{\rm T}^{\rm eff}=p+(\rho/2\lambda _{\rm b})(\rho+2p)+(\widetilde{U}-\widetilde{P})$. Thus structurally, this is identical to the result presented in the previous section with $d=4$ with $\rho$, $p$ and $p_{\rm T}$ replaced by $\rho_{\rm eff}$, $p_{\rm eff}$ and $p_{\rm T}^{\rm eff}$ respectively. Thus one would naively suggest that the bound on the photon circular orbit, namely $r_{\rm ph}\leq 3\mathcal{M}$ should remain valid, where $\mathcal{M}$ is the ADM mass of the spacetime.

However, the validity of the result derived in \ref{Ph_GR_CC_HD} requires a series of assumptions to hold true and since the extra piece originating from the extra dimensions is not required to satisfy the energy conditions, the bound may get violated. Let us then discuss, which of the assumptions presented in the earlier derivation may get violated. First of all, the solution for $e^{-\lambda}$ will now read,
\begin{align}\label{Eq_Ph_23}
e^{-\lambda}=1-\frac{2m(r)}{r}-\frac{\Lambda_{4}}{3}r^{2};\qquad m(r)=M_{\rm H}+4\pi \int _{r_{H}}^{r}dr' \rho_{\rm eff}(r')r'^{d-2}
\end{align}
which will be assumed to vanish at some radius $r=r_{\rm H}$, which is the black hole horizon and also at $r=r_{\rm C}$, the cosmological horizon. Here, $M_{\rm H}$ is the mass of the black hole. As we subtract \ref{Eq_Ph_22a} and \ref{Eq_Ph_22b} it immediately follows that $\rho_{\rm eff}(r_{\rm H})+p_{\rm eff}(r_{\rm H})=0$, since $e^{-\lambda}(\nu'+\lambda')$ vanishes at the horizon. Following which, one may argue that $p_{\rm eff}(r_{\rm H})<0$, if the effective density at the horizon is a positive definite quantity. For the matter energy density $\rho$, this is certainly true, however for the contribution from the bulk Weyl tensor, similar results cannot be accounted for, i.e., $\widetilde{U}$ can be negative and hence the total effective energy density $\rho_{\rm eff}$ need not be a positive definite quantity. Thus if the matter contribution is larger than the bulk contribution, $\rho_{\rm eff}$ is positive definite and the previous bound on the photon circular orbit still applies. On the other hand, if the bulk contribution dominates, then $\rho_{\rm eff}$ is negative, which would imply that $p_{\rm eff}(r_{\rm H})>0$, in contrast to the previous scenario. 

Let us proceed further to understand how this behaviour of the effective pressure will affect the bound on the photon circular orbit. First of all, the photon circular orbit on the equatorial plane is a solution to the algebraic equation, $r\nu'=2$, which on using \ref{Eq_Ph_22b}, yields the following algebraic equation,
\begin{align}\label{Eq_Ph_N06}
\left(8\pi p_{\rm eff}r^{2}-\Lambda_{4}r^{2}\right)+\left(1-e^{-\lambda}\right)=2e^{-\lambda}~,
\end{align}
which is independent of the metric degree of freedom $\nu(r)$ and is dependent only on $\lambda(r)$, the matter variables and the dark radiation and pressure inherited from the bulk spacetime. Following the situation in general relativity, let us define the following quantity,
\begin{align}\label{Eq_Ph_N07}
\mathcal{N}_{\rm brane}(r)\equiv -8\pi p_{\rm eff}r^{2}+\Lambda_{4}r^{2}-1+3e^{-\lambda}~.
\end{align}
As evident from \ref{Eq_Ph_N06}, on the photon circular orbit $r_{\rm ph}$, we have $\mathcal{N}_{\rm brane}(r_{\rm ph})=0$. Using the solution for $e^{-\lambda}$ from \ref{Eq_Ph_23}, the function $\mathcal{N}_{\rm brane}(r)$ yields,
\begin{align}\label{Eq_Ph_N12}
\mathcal{N}_{\rm brane}(r)=-8\pi p_{\rm eff}r^{2}+\Lambda_{4}r^{2}-1+3\left(1-\frac{2m(r)}{r}-\frac{\Lambda}{3}r^{2}\right)
=2-\frac{6m(r)}{r}-8\pi p_{\rm eff}r^{2}~,
\end{align}
which is independent of the brane cosmological constant $\Lambda_{4}$. It is further assumed that both the matter energy density $\rho(r)$ and the pressure $p(r)$ decays sufficiently fast, so that, $pr^{2}\rightarrow 0$. In addition, the dark radiation and dark pressure terms are also assumed to decay sufficiently fast, such that $m(r)\rightarrow \textrm{constant}$ as $r\rightarrow \infty$. Thus from \ref{Eq_Ph_N12} it immediately follows that,
\begin{align}\label{Eq_Ph_N13}
\mathcal{N}_{\rm brane}(r\rightarrow \infty)=2~.
\end{align}
Note that this asymptotic limit is independent of the presence of higher dimensions. Also note that the result $\lambda'e^{-\lambda}<0$ at the black hole horizon will hold irrespective of the sign of $\rho_{\rm eff}$ and hence $\mathcal{N}_{\rm brane}(r_{\rm H})<0$ will follow. Further, the conservation of energy momentum tensor will follow an identical route to the one adopted in \ref{Ph_GR_CC_HD}, ultimately resulting into, 
\begin{align}\label{Eq_Ph_24}
P_{\rm eff}'(r)=\frac{r^{3}e^{\lambda}}{2}\Big[\left(p_{\rm eff}+\rho_{\rm eff}\right)\mathcal{N}+2e^{-\lambda}\left\{-\rho_{\rm eff}+p_{\rm eff}+2p^{\rm eff}_{\rm T}\right\} \Big]~,
\end{align}
where, $P_{\rm eff}\equiv r^{4}p_{\rm eff}$. Intriguingly, the contributions from the bulk Weyl tensor is traceless and hence only the matter energy density and pressure contributes in \ref{Eq_Ph_24}. Then even on the brane, we have the following relation, $-\rho_{\rm eff}+p_{\rm eff}+2p^{\rm eff}_{\rm T}<0$ \cite{Hod:2020pim}. Hence, $P_{\rm eff}'(r_{\rm ph})<0$ and so is the case on the horizon as well. Thus \ref{Eq_Ph_18} holds true even in this situation. However, since $p_{\rm eff}(r_{\rm H})$ can be positive, this means $p_{\rm eff}(r_{\rm ph})$ can also be positive. The fact that $p(r)$ decreases with radial distance is taken care of by the result that $p(r)$ must depend on inverse powers of the radial distance $r$, which follows from the asymptotic fall-off conditions. This in turn implies, $p(r_{\rm ph})\geq 0$ and hence the bound on the photon circular orbit now translates into, $r_{\rm ph}\geq 3m(r_{\rm ph})>3M_{\rm H}$, where $M_{\rm H}$ is the mass of the black hole and is less than the ADM mass, which is obtained as the limiting procedure, $m(r\rightarrow \infty)$. Therefore, when the contribution from the bulk spacetime dominates over the matter sector and is negative, it follows that the photon sphere has a lower bound, rather than an upper bound, given by,  
\begin{align}\label{Eq_Ph_25}
r_{\rm ph}\geq 3M_{\rm H}~.
\end{align}
This is in complete contrast with the previous scenario, and also from the case in which matter energy density on the brane dominates over the bulk contributions, where instead of an upper bound we have arrived at a lower bound on the photon circular orbit (see also \cite{Guo:2020zmf}). To see that this is indeed the situation in case of a braneworld black hole, consider the solution presented in \cite{Dadhich:2000am}, where both $\rho$ and $p$ are vanishing while $\widetilde{U}\sim -\widetilde{U}_{0}/r^{4}$ and $\widetilde{P}=-2\widetilde{U}$. Thus the effective energy density $\rho_{\rm eff}=3\widetilde{U}$ is negative, while the pressure $p_{\rm eff}=\widetilde{U}+2\widetilde{P}$ is positive. The black hole solution for the above configuration has the following structure,
\begin{align}\label{Eq_Ph_26}
e^{\nu}=e^{-\lambda}=1-\frac{2M}{r}-\frac{q}{r^{2}}-\frac{\Lambda_{4}}{3}r^{2}~,
\end{align}
with the photon sphere being located at the solution of the algebraic equation, 
\begin{align}\label{Eq_Ph_27}
2\left(1-\frac{2M}{r}-\frac{q}{r^{2}}-\frac{\Lambda_{4}}{3}r^{2}\right)=r\left(\frac{2M}{r^{2}}+\frac{2q}{r^{3}}-2\frac{\Lambda_{4}}{3}r\right)~,
\end{align}
which is also equivalent to the following one: $r^{2}-3Mr-2q=0$. This has the following solution for the radius of the photon sphere, $r_{\rm ph}=(1/2)(3M+\sqrt{9M^{2}+8q})>3M$, consistent with the bound derived in \ref{Eq_Ph_25}. We must point out that since the function $\widetilde{U}$, inherited from the higher dimension is not necessarily a positive definite quantity, the Hawking-Geroch mass is not guaranteed to be positive, unlike the general relativistic scenario, discussed above. This is why we have not pursued the alternative approach to arrive at the bound on the photon circular orbit in the present context, rather, we leave it for a future work.  

On the other hand, as emphasized earlier, if a matter field is present on the brane, then there will be a competition between the term $\widetilde{U}$ and the terms involving $\rho$ and $\rho^{2}$. If the contribution from the bulk Weyl tensor $\widetilde{U}$ dominates over and above the contribution from the brane matter, the photon circular orbit will again satisfy \ref{Eq_Ph_25}. While, if the matter contributions from the brane dominate, the photon circular orbit satisfies \ref{Eq_Ph_20}. A direct illustration of the above result can be achieved by considering Maxwell field on the brane. In which case the metric elements take the structure of \ref{Eq_Ph_26} with $-q$ replaced by $Q^{2}-q$, where $Q^{2}$ corresponds to Maxwell charge\footnote{There will be some additional terms $\sim \mathcal{O}(\ell^{2})$, where $\ell$ is related to the size of the extra dimension. Since the extra dimension must be compact, it follows that $(\ell/r_{\rm ph})\ll 1$ and hence effect of these terms can be ignored.}. Thus the location of photon circular orbit corresponds to $r_{\rm ph}=(1/2)(3M+\sqrt{9M^{2}-8Q^{2}+8q})$. As evident, for $Q^{2}>q$, i.e., when Maxwell charge dominates it yields $r_{\rm ph}\leq 3M$, while for $Q^{2}<q$, the contribution from bulk Weyl tensor dominates and hence $r_{\rm ph}\geq 3M$. This is in complete consonance with our earlier discussion. Thus in the braneworld scenario the bound on the photon circular orbit do exists, but whether it is an upper bound or a lower bound depends on whether the effect from bulk Weyl tensor dominates over and above the brane matter distribution or not.
\section{Bound on photon circular orbit in pure Lovelock gravity}\label{Ph_pureLovelock}

Having discussed the effect of extra dimensions on the location of the photon sphere, both within the purview of general relativity as well as for effective four dimensional theories, we will discuss the corresponding situation within the context of higher curvature theories of gravity in this section. As one of the most important sub-class of higher curvature theories, we will consider possible bound on the photon circular orbit in the context of pure Lovelock gravity. As we have seen in the earlier sections, presence of the cosmological constant has no effect on the photon circular orbits and hence we will exclusively work with the asymptotically flat scenario. For that purpose, once again, we start with the static and spherically symmetric metric ansatz as presented in \ref{Eq_Ph_01}. The field equations for pure Lovelock gravity, involving the two unknowns, namely $\nu(r)$ and $\lambda(r)$, in the presence of perfect fluid takes the following form,
\begin{align}
8\pi \rho(r)&=\frac{\left(1-e^{-\lambda}\right)^{N-1}}{2^{N-1}r^{2N}}\Bigg[rN\lambda 'e^{-\lambda}+\left(d-2N-1\right)\left(1-e^{-\lambda}\right)\Bigg]~,
\label{Eq_Ph_28a}
\\
8\pi p(r)&=\frac{\left(1-e^{-\lambda}\right)^{N-1}}{2^{N-1}r^{2N}}\Bigg[rN\nu'e^{-\lambda}-\left(d-2N-1\right)\left(1-e^{-\lambda}\right)\Bigg]~.
\label{Eq_Ph_28b}
\end{align}
Here, $d$ stands for the spacetime dimensions and $N$ corresponds to the order of the pure Lovelock polynomial, e.g., $N=1$ corresponds to general relativity, $N=2$ corresponds to the Gauss-Bonnet term and so on. We further assume that there exist a radius $r=r_{\rm H}$, such that, $e^{-\lambda}(r_{\rm H})=0$, signalling the presence of a horizon at this radius. From these results it follows that $e^{-\lambda}(\lambda'+\nu')=0$ at $r=r_{\rm H}$, while from the previous discussions we observe that $e^{-\lambda}\lambda'<0$ on the horizon as well. These two conditions will form the main ingredient of this section. Proceeding further, we observe that similar to the previous scenarios considered here, as the above two field equations, presented in \ref{Eq_Ph_28a} and \ref{Eq_Ph_28b} are being added up, on the black hole horizon $r_{\rm H}$, the following relation holds,
\begin{align}\label{Eq_Ph_29}
p(r_{\rm H})+\rho(r_{\rm H})=0~.
\end{align}
Thus for normal matter fields, satisfying $\rho(r)>0$, for all possible choices of $r$, it follows that $p(r_{\rm H})<0$. This will turn out to be a useful relation in deriving the bound on the photon circular orbit. Determination of the photon circular orbit involves two steps, first, one must solve for the metric coefficient $\nu'$ starting from the above gravitational field equations, in particular \ref{Eq_Ph_28b}, and then the corresponding expression must be substituted in the algebraic relation, given by $r\nu'=2$. This procedure, in the present context, results into the following algebraic equation, 
\begin{align}\label{Eq_Ph_31}
2e^{-\lambda}=\frac{(d-2N-1)(1-e^{-\lambda})}{N}+\frac{8\pi}{N}\frac{2^{N-1}r^{2N}p(r)}{(1-e^{-\lambda})^{N-1}}~.
\end{align}
This prompts one to define, in analogy with the corresponding general relativistic counter part, the following quantity
\begin{align}\label{Eq_Ph_32}
\mathcal{N}_{\rm lovelock}(r)=(d-1)e^{-\lambda}-(d-2N-1)-8\pi \frac{2^{N-1}r^{2N}p(r)}{(1-e^{-\lambda})^{N-1}}~,
\end{align}
which, by definition vanishes at the photon circular orbit, located at $r=r_{\rm ph}$. To understand the behaviour of the function $\mathcal{N}_{\rm lovelock}(r)$ at the black hole horizon, it is desirable to write down the expression for $\lambda'$ on $r=r_{\rm H}$. Starting from the gravitational field equation presented in \ref{Eq_Ph_28a}, we obtain,
\begin{align}\label{Eq_Ph_33}
r_{\rm H}N\lambda'(r_{\rm H})e^{-\lambda\left(r_{\rm H}\right)}=-(d-2N-1)+8\pi \times 2^{N-1}r_{\rm H}^{2N}\rho(r_{\rm H})~.
\end{align}
Since, from our earlier discussion it follows that $\lambda'(r_{\rm H})e^{-\lambda(r_{\rm H})}<0$, it is immediate that the term on the right hand side of \ref{Eq_Ph_33} is negative, when evaluated at the location of the horizon. Therefore the quantity $\mathcal{N}_{\rm lovelock}(r_{\rm H})$, becomes,
\begin{align}\label{Eq_Ph_34}
\mathcal{N}_{\rm lovelock}(r_{\rm H})=-(d-2N-1)-8\pi \times 2^{N-1}r_{\rm H}^{2N}p(r_{\rm H})=r_{\rm H}N\lambda'\left(r_{\rm H}\right)
e^{-\lambda\left(r_{\rm H}\right)}<0~.
\end{align}
The last bit follows from the result $\rho(r_{\rm H})=-p(r_{\rm H})$, presented in \ref{Eq_Ph_29}. Also in the asymptotic limit, for pure lovelock theories, the appropriate fall-off conditions for the components of the matter energy momentum tensor are such that: $p(r)r^{2N}\rightarrow 0$ and $e^{-\lambda}\rightarrow 1$. Thus, we obtain the asymptotic form of the function $\mathcal{N}_{\rm lovelock}(r)$ to read, 
\begin{align}\label{Eq_Ph_35}
\mathcal{N}_{\rm lovelock}(r\rightarrow \infty)=(d-1)-(d-2N-1)=2N~.
\end{align}
As evident, for pure Lovelock theory of order $N$ the asymptotic value of the quantity $\mathcal{N}_{\rm lovelock}(r)$ is dependent on the order of the Lovelock polynomial. For general relativity, which has $N=1$, the asymptotic value of $\mathcal{N}_{\rm lovelock}(r)$ is $2$, consistent with earlier observations.

To proceed further, we need to solve for the metric coefficient $e^{-\lambda}$. This can be done by first writing down the differential equation for $\lambda(r)$, presented in \ref{Eq_Ph_28a}, as a first order differential equation, whose integration yields,
\begin{align}\label{Eq_Ph_37}
e^{-\lambda}=1-2\left(\frac{m(r)}{r^{d-2N-1}}\right)^{1/N};\qquad m(r)=M_{\rm H}+4\pi \int _{r_{\rm H}}^{r}dr'~\rho(r')r'^{d-2}~,
\end{align}
where, $M_{\rm H}=(r_{\rm H}^{d-2N-1}/2^{N})$ is the mass of the black hole spacetime and is less than the ADM mass $\mathcal{M}$, which contains contribution from the matter energy density $\rho$ as well. The final ingredient necessary for the rest of the computation is the conservation of the matter energy momentum tensor, which does not depend on the gravity theory under consideration, and it reads
\begin{align}\label{Eq_Ph_38}
p'(r)+\frac{\nu'}{2}\left(p+\rho\right)+\left(\frac{d-2}{r}\right)\left(p-p_{\rm T}\right)=0~.
\end{align}
One can solve for $\nu'$ from the above equation, which when equated to the corresponding expression from the gravitational field equations, namely from \ref{Eq_Ph_28b}, results a differential equation for the radial pressure $p(r)$. This differential equation can be further simplified by introducing the quantity $\mathcal{N}_{\rm lovelock}(r)$, which ultimately results into,
\begin{align}\label{Eq_Ph_40}
p'(r)&=\frac{e^{\lambda}}{2Nr}\Big[\left(p+\rho\right)\mathcal{N}+2Ne^{-\lambda}\left\{-\rho+p+(d-2)p_{T}\right\}-2dNe^{-\lambda}p(r)\Big]~.
\end{align}
Following our previous considerations, we can define another quantity, $P(r)\equiv r^{d}p(r)$, where $d$ stands for the spacetime dimensions. Then the differential equation satisfied by $P(r)$ takes the following form, 
\begin{align}\label{Eq_Ph_41}
P'(r)&=r^{d}p'(r)+dr^{d-1}p(r)
\nonumber
\\
&=\frac{r^{d-1}e^{\lambda}}{2N}\Big[\left(p+\rho\right)\mathcal{N}+2Ne^{-\lambda}\left\{-\rho+p+(d-2)p_{T}\right\}
-2dNe^{-\lambda}p(r)\Big]+dr^{d-1}p(r)
\nonumber
\\
&=\frac{r^{d-1}e^{\lambda}}{2N}\Big[\left(p+\rho\right)\mathcal{N}+2Ne^{-\lambda}\left\{-\rho+p+(d-2)p_{T}\right\}\Big]~.
\end{align}
When evaluated at the location of the black hole horizon, $r=r_{\rm H}$, this differential equation for $P(r)$ yields the following inequality, 
\begin{align}\label{Eq_Ph_42}
P'(r_{\rm H})=\frac{r_{\rm H}^{d-1}e^{\lambda(r_{\rm H})}}{2N}\Big[\left\{p(r_{\rm H})+\rho(r_{\rm H})\right\}\mathcal{N}(r_{\rm H})+2Ne^{-\lambda(r_{\rm H})}\left\{-\rho(r_{\rm H})+p(r_{\rm H})+(d-2)p_{T}(r_{\rm H})\right\}\Big]<0~,
\end{align}
which follows from the result that $e^{-\lambda}$ vanishes at $r=r_{\rm H}$ and $\mathcal{N}_{\rm lovelock}(r_{\rm H})<0$ (see \ref{Eq_Ph_34} for a derivation). Further, note that even when evaluated at $r=r_{\rm ph}$, the object $P'(r)$ is negative. This is because, $\mathcal{N}(r_{\rm ph})=0$ by definition and the trace of the energy-momentum tensor $-\rho+p+(d-2)p_{\rm T}$ is also assumed to be negative. Thus, we finally arrive at the following condition, 
\begin{align}\label{Eq_Ph_43}
P'(r_{\rm H}\leq r\leq r_{\rm ph})<0~,
\end{align}
for black holes in pure Lovelock theories of gravity. This means that the quantity $P(r)$ and hence $p(r)$ decreases as the radius is increasing from the black hole horizon to the photon circular orbit. Since $p(r_{\rm H})<0$, it immediately follows that $p(r_{\rm ph})<0$ as well. Using this result along with $\mathcal{N}(r_{\rm ph})=0$, yields,
\begin{align}\label{Eq_Ph_44}
\left(d-1\right)e^{-\lambda(r_{\rm ph})}-\left(d-2N-1\right)=8\pi 2^{N-1}\frac{r^{2N}p(r_{\rm ph})}{(1-e^{-\lambda})^{N-1}}<0
\end{align}
Substitution of the corresponding expression for $e^{-\lambda}$ from \ref{Eq_Ph_37} results into the following upper bound on the location of the photon circular orbit,
\begin{align}\label{Eq_Ph_46}
r_{\rm ph}\leq \Big[\left(\frac{d-1}{2N}\right)^{N}2\mathcal{M}\Big]^{1/(d-2N-1)}~.
\end{align}
As evident, for $d=4$ and $N=1$, the right hand side becomes $3\mathcal{M}$, while for arbitrary $d$ with $N=1$, we get back our previous result, presented in \ref{Eq_Ph_20}. Hence the general relativistic limit is reproduced for any spacetime dimensions. Thus the above provides the upper bound on the location of the photon circular orbit $r_{\rm ph}$ for any pure Lovelock theory of order $N$, in any spacetime dimension $d$. 
\section{Bound on photon circular orbit in Einstein-Gauss Bonnet gravity}\label{Ph_EGB}

Having discussed the case of pure Lovelock gravity in the previous section, we will now take up the case of general Lovelock theories. As a warm up to that direction, we present a brief analysis of five dimensional Einstein-Gauss-Bonnet gravity in the present section and the associated bound on the location of the photon circular orbit. To start with, we write down the gravitational field equations in the Einstein-Gauss-Bonnet gravity, which takes the following form, 
\begin{align}
8\pi r^{2}\rho (r)&=r\lambda 'e^{-\lambda}+2\left(1-e^{-\lambda}\right)
+\alpha \frac{(1-e^{-\lambda})}{r^{2}}\times 2r\lambda 'e^{-\lambda}~,
\label{egb_ph_01}
\\
8\pi r^{2}p(r)&=r\nu'e^{-\lambda}-2\left(1-e^{-\lambda}\right)+\alpha \frac{(1-e^{-\lambda})}{r^{2}}\times 2r\nu 'e^{-\lambda}~.
\label{egb_ph_02}
\end{align}
Here, $\alpha$ is the Gauss-Bonnet coupling, which is the coefficient of the $(R^{2}-4R_{ab}R^{ab}+R_{abcd}R^{abcd})$ term in the five-dimensional gravitational Lagrangian. As usual, the algebraic equation $e^{-\lambda}(r)=0$, defines the location of the horizon $r_{\rm H}$, while, our previous analysis guarantees that  $\lambda'(r_{\rm H})e^{-\lambda(r_{\rm H})}<0$. Then from the addition of the above field equations, it follows that $\rho(r_{\rm H})+p(r_{\rm H})=0$, owing to the fact that $\lambda '(r_{\rm H})+\nu'(r_{\rm H})$ is finite, but $e^{-\lambda(r_{\rm H})}$ is vanishing. Thus for positive matter energy density, it follows that the pressure on the horizon must be negative. This will be a crucial result in obtaining the bound on the photon circular orbit.

The equation involving the unknown metric coefficient $\lambda(r)$, from \ref{egb_ph_01}, can be expressed as a simple first order differential equation, whose integration yields the following solution for $1-e^{-\lambda(r)}$,
\begin{align}
1-e^{-\lambda}=-\frac{r^{2}}{2\alpha}+\frac{r^{2}}{2\alpha}\sqrt{1+\frac{8\alpha m(r)}{r^{4}}}~;
\qquad
m(r)=M_{\rm H}+4\pi \int_{r_{\rm H}}^{r}dr'\rho(r')r'^{3}~.
\end{align}
Here, $M_{\rm H}$ is the mass of the black hole and the above solution is so chosen, such that the spacetime is asymptotically flat. The pressure equation, i.e., \ref{egb_ph_02}, on the other hand, can be solved to obtain an expression for $\nu'$. This when used in the algebraic expression for the photon circular orbit. i.e., in the relation $r\nu'=2$, we obtain,
\begin{align}
2e^{-\lambda(r_{\rm ph})}=\frac{8\pi r_{\rm ph}^{2}p(r_{\rm ph})+2\left(1-e^{-\lambda(r_{\rm ph})}\right)}{1+\frac{2\alpha}{r_{\rm ph}^{2}}\left(1-e^{-\lambda(r_{\rm ph})}\right)}~.
\end{align}
This suggests to define the following function,
\begin{align}
\mathcal{N}_{\rm EGB}(r)\equiv 4e^{-\lambda}-2-8\pi r^{2}p(r)+\frac{4\alpha}{r^{2}}e^{-\lambda}\left(1-e^{-\lambda}\right)~.
\end{align}
Motivated by the results presented in the previous sections, expressions for the quantity $\mathcal{N}_{\rm EGB}$ on the horizon, on the photon circular orbit and also the asymptotic value of $\mathcal{N}_{\rm EGB}$ are of importance. These values are given by,
\begin{align}
\mathcal{N}_{\rm EGB}(r_{\rm ph})&=0~;\qquad  \mathcal{N}_{\rm EGB}(r\rightarrow \infty)=2~,
\\
\mathcal{N}_{\rm EGB}(r_{\rm H})&=-2-8\pi r_{\rm H}^{2}p(r_{\rm H})=-2+8\pi r_{\rm H}^{2}\rho(r_{\rm H})=\left[r_{\rm H}+\frac{2\alpha}{r_{\rm H}}\right]\lambda'(r_{\rm H})e^{-\lambda(r_{\rm H})}<0~.
\end{align}
In the five dimensional static and spherically symmetric spacetime, which is a solution of the Einstein-Gauss-Bonnet theory, the matter energy momentum tensor conservation takes the following form, 
\begin{align}
p'(r)+\frac{\nu'}{2}\left(p+\rho \right)+\frac{3}{r}\left(p-p_{\rm T}\right)=0~,
\end{align}
where, $p_{\rm T}$ corresponds to the transverse pressure. Substituting the value for $\nu'$ from \ref{egb_ph_02}, we obtain the following expression for $(dp/dr)$, after simplifications,  
\begin{align}
p'(r)&=\frac{e^{\lambda}}{2r}\frac{1}{1+\frac{2\alpha}{r^{2}}(1-e^{-\lambda})}
\Bigg[\left(p+\rho\right)\mathcal{N}_{\rm EGB}+2e^{-\lambda}\left(-\rho+p+3p_{\rm T}\right)
\left\{1+\frac{2\alpha}{r^{2}}(1-e^{-\lambda})\right\}
\nonumber
\\
&\hskip 4 cm -10pe^{-\lambda}\left\{1+\frac{2\alpha}{r^{2}}(1-e^{-\lambda})\right\}\Bigg]~.
\end{align}
Taking a cue from the earlier considerations, we may now define a rescaled radial pressure, $P(r)\equiv r^{5}p(r)$, whose derivative becomes, 
\begin{align}
P'(r)&=r^{5}p'(r)+5r^{4}p(r)
\nonumber
\\
&=\frac{r^{4}e^{\lambda}}{2}\frac{1}{1+\frac{2\alpha}{r^{2}}(1-e^{-\lambda})}\Bigg[\left(p+\rho\right)\mathcal{N}_{\rm EGB}+2e^{-\lambda}\left(-\rho+p+3p_{\rm T}\right)\left\{1+\frac{2\alpha}{r^{2}}(1-e^{-\lambda})\right\}\Bigg]~.
\end{align}
As evident, since $\mathcal{N}_{\rm EGB}(r_{\rm ph})=0$ and $\mathcal{N}_{\rm EGB}(r_{\rm H})<0$, it follows that, $P'(r_{\rm H}\leq r\leq r_{\rm ph})<0$. Thus it follows that $p(r_{\rm ph})<0$ as well, since $p(r_{\rm H})$ must be negative and the radial pressure $p(r)$ is a monotonically decreasing function within the range $r_{\rm H}\leq r\leq r_{\rm ph}$. Thus we obtain, from the fact that $\mathcal{N}_{\rm EGB}(r_{\rm ph})=0$, that on the photon sphere the following algebraic relation holds true,
\begin{align}
2e^{-\lambda(r_{\rm ph})}+\frac{2\alpha}{r_{\rm ph}^{2}}e^{-\lambda(r_{\rm ph})}\left(1-e^{-\lambda(r_{\rm ph})}\right)\leq 1~.
\end{align}
The above being an inequality on the quadratic function of $e^{-\lambda(r_{\rm ph})}$, demands that on the photon sphere the following inequality must hold true,
\begin{align}\label{boundegb}
e^{-\lambda(r_{\rm ph})}\leq \frac{1}{2}\left[1+\frac{r_{\rm ph}^{2}}{2\alpha}-\sqrt{1+\frac{r_{\rm ph}^{4}}{\alpha ^{2}}}\right]~.
\end{align}
Substituting the solution for $e^{-\lambda(r_{\rm ph})}$ in the above inequality, we obtain the following bound on the radius of the photon circular orbit, 
\begin{align}
r_{\rm ph}\leq \left[16m(r_{\rm ph})^{2}-8\alpha m(r_{\rm ph})\right]^{1/4}\leq 2\sqrt{\mathcal{M}}~.
\end{align}
Note that the substitution, $m\rightarrow (Gm/6\pi^{2})$ will make the above inequality coincide\footnote{Note that the mass appearing in the spacetime metric is always scaled by the gravitational constant, thus $m\rightarrow Gm$, when the gravitational constant is restored. However, often in the literature the mass is also scaled by the inverse of the factor $(d-2)S_{d-2}$, with one caveat. In the Schwarzschild limit one obtains the gravitational potential to be, $(2GM/2S_{2})r^{-1}=(GM/4\pi r)$. Due to this unusual scaling of the mass we have avoided this factor in the present analysis. Thus to match with the literature, all the masses may be scaled by $(G/(d-2)S_{d-2})$. In five dimensions, this scaling becomes $(G/6\pi^{2})$, as $S_{3}=2\pi^{2}$.} with the one derived in \cite{Gallo:2015bda}. Since the ADM mass $\mathcal{M}$ of the spacetime involves contribution from the matter fields falling within the black hole horizon, through the term $\rho(r)$, it follows that $m(r_{\rm ph})\leq \mathcal{M}$. This has been used to arrive at the final inequality, presented above. Thus we have demonstrated the versatility of the method depicted here, as it yielded the desired upper bound on the location of the photon circular orbit, in terms of the ADM mass of the spacetime. 

\section{Bound on the photon circular orbit in general Lovelock gravity}\label{Ph_genLovelock}

After discussing the bound on the photon circular orbit in Einstein-Gauss-Bonnet gravity, let us determine the corresponding bound for general Lovelock Lagrangian, where along with Einstein Lagrangian several other higher order Lovelock terms appear. Let us work in $d$-dimensions, involving $N$ Lovelock polynomials, with the maximum order of the Lovelock polynomial being, $N_{\rm max}=(d-2)/2$. This is because, for Lovelock theories involving $N>N_{\rm max}$, there are no propagating gravitational degrees of freedom. In particular, for $N=(d/2)$, the Lovelock polynomial becomes a total derivative. This situation can be compared to that of general relativity, which has dynamics in four dimensions, but is devoid of dynamics in three and two dimensions. 

As in the previous scenarios, in the context of general Lovelock gravity as well, the first step in deriving the bound on the photon circular orbit corresponds to writing down the temporal and the radial components of the gravitational field equations, which take the following form \cite{Dadhich:2016fku},
\begin{align}
\sum _{m}\hat{\alpha}_{m}\frac{(1-e^{-\lambda})^{m-1}}{r^{2(m-1)}}
\left[mr\lambda'e^{-\lambda}+(d-2m-1)(1-e^{-\lambda}) \right]&=8\pi r^{2}\rho~,
\label{genL_eq_01}
\\
\sum _{m}\hat{\alpha}_{m}\frac{(1-e^{-\lambda})^{m-1}}{r^{2(m-1)}}
\left[mr\nu'e^{-\lambda}-(d-2m-1)(1-e^{-\lambda}) \right]&=8\pi r^{2}p~.
\label{genL_eq_02}
\end{align}
where, $\hat{\alpha}_{m}\equiv (1/2)\{(d-2)!/(d-2m-1)!\}\alpha _{m}$, with $\alpha _{m}$ being the coupling constant appearing in the $m$th order Lovelock Lagrangian. Also note that, the summation in the above field equations must run from $m=1$ to $m=N_{\rm max}$. Since $e^{-\lambda}$ vanishes on the event horizon located at $r=r_{\rm H}$, addition of \ref{genL_eq_01} and \ref{genL_eq_02} yields, 
\begin{align}
8\pi r_{\rm H}^{2}[\rho(r_{\rm H})+p(r_{\rm H})]=0~,
\end{align}
which suggests that the pressure at the horizon must be negative, if the matter field satisfies the weak energy condition, i.e., $\rho>0$. Further we can determine an analytic expression for $\nu'$, starting from \ref{genL_eq_02}. This, when used in association with the fact that on the photon circular orbit, $r\nu'=2$, it follows that,
\begin{align}
2e^{-\lambda(r_{\rm ph})}\sum _{m}m\hat{\alpha}_{m}\frac{(1-e^{-\lambda(r_{\rm ph})})^{m-1}}{r_{\rm ph}^{2(m-1)}}
=8\pi r_{\rm ph}^{2}p(r_{\rm ph})+\sum _{m}\hat{\alpha}_{m}(d-2m-1)\frac{(1-e^{-\lambda(r_{\rm ph})})^{m}}{r_{\rm ph}^{2(m-1)}}~.
\end{align}
This prompts one to define the following object,
\begin{align}
\mathcal{N}_{\rm gen}(r)=2e^{-\lambda}\sum _{m}m\hat{\alpha}_{m}\frac{(1-e^{-\lambda})^{m-1}}{r^{2(m-1)}}-8\pi r^{2}p-\sum _{m}\hat{\alpha}_{m}(d-2m-1)\frac{(1-e^{-\lambda})^{m}}{r^{2(m-1)}}~.
\end{align}
As in the case of Einstein-Gauss-Bonnet gravity, for general Lovelock theory as well, it follows that $\mathcal{N}_{\rm gen}(r_{\rm ph})=0$ and also $\mathcal{N}_{\rm gen}(r_{\rm H})<0$. Further in the asymptotic limit, if we assume the solution to be asymptotically flat then, only the $m=1$ term in the above series will survive, as $e^{-\lambda}\rightarrow 1$ as $r\rightarrow \infty$. Thus even in this case $\mathcal{N}_{\rm gen}(r\rightarrow \infty)=2$. 

To proceed further, we consider the conservation equation for the matter energy momentum tensor, which in $d$ spacetime dimensions has been presented in \ref{Eq_Ph_14}. As usual, this conservation equation can be rewritten using the expression for $\nu'$ from \ref{genL_eq_02}, such that,
\begin{align}
p'&=\frac{e^{\lambda}}{2r}\frac{1}{\left\{\sum_{m}m\hat{\alpha}_{m}\frac{(1-e^{-\lambda})^{m-1}}{r^{2(m-1)}}\right\}}
\Bigg[(\rho+p)\mathcal{N}_{\rm gen}+2e^{-\lambda}\left\{-\rho+p+(d-2)p_{\rm T}\right\}\sum _{m}m\hat{\alpha}_{m}\frac{(1-e^{-\lambda})^{m-1}}{r^{2(m-1)}}
\nonumber
\\
&\hskip 6 cm -2dpe^{-\lambda}\sum _{m}m\hat{\alpha}_{m}\frac{(1-e^{-\lambda})^{m-1}}{r^{2(m-1)}}\Bigg]~.
\end{align}
In this case the rescaled radial pressure, defined as $P(r)\equiv r^{d}p(r)$, satisfies the following first order differential equation,
\begin{align}
P'&=r^{d}p'+dr^{d-1}p
\nonumber
\\
&=\frac{e^{\lambda}}{2}\frac{r^{d-1}}{\left\{\sum _{m}m\hat{\alpha}_{m}\frac{(1-e^{-\lambda})^{m-1}}{r^{2(m-1)}}\right\}}
\Bigg[(\rho+p)\mathcal{N}_{\rm gen}+2e^{-\lambda}\left\{-\rho+p+(d-2)p_{T}\right\}\sum _{m}m\hat{\alpha}_{m}\frac{(1-e^{-\lambda})^{m-1}}{r^{2(m-1)}}\Bigg]~.
\end{align}
It is evident from the results, $\mathcal{N}_{\rm gen}(r_{\rm ph})=0$ and $\mathcal{N}_{\rm gen}(r_{\rm H})<0$, that $P'(r)$ is certainly negative within the region bounded by the horizon and the photon circular orbit. Since, $p(r_{\rm H})$ is negative, it further follows that $p(r_{\rm ph})\leq 0$ as well. Thus from the definition of $\mathcal{N}_{\rm gen}$ and the result that $\mathcal{N}_{\rm gen}(r_{\rm ph})=0$, it follows that, 
\begin{align}\label{condition_extra}
\sum _{m=1}^{N_{\rm max}}\hat{\alpha}_{m}\frac{(1-e^{-\lambda(r_{\rm ph})})^{m-1}}{r_{\rm ph}^{2(m-1)}}\left[2me^{-\lambda(r_{\rm ph})}-(d-2m-1)(1-e^{-\lambda(r_{\rm ph})}) \right]\leq 0~.
\end{align}
Here, the coupling constants $\hat{\alpha}_{m}$'s are assumed to be positive. Also, $e^{-\lambda}$ vanishes on the horizon and reaches unity asymptotically, such that for any intermediate radius, e.g., at $r=r_{\rm ph}$, $e^{-\lambda}$ is positive and less than unity, such that $(1-e^{-\lambda(r_{\rm ph})})>0$. Thus the quantity within bracket in \ref{condition_extra} will determine the fate of the above inequality. Note that, if the above inequality holds for $N=N_{\rm max}$, i.e., if we impose the condition,
\begin{align}\label{maxineq}
2N_{\rm max}e^{-\lambda(r_{\rm ph})}-(d-2N_{\rm max}-1)(1-e^{-\lambda(r_{\rm ph})}) \leq 0~.
\end{align}
Then it follows that, for any $N=(N_{\rm max}-n)<N_{\rm max}$ (with integer $n$), we have, 
\begin{align}
2Ne^{-\lambda(r_{\rm ph})}&-(d-2N-1)(1-e^{-\lambda(r_{\rm ph})})
\nonumber
\\
&=2\left(N_{\rm max}-n\right)e^{-\lambda(r_{\rm ph})}-[d-2\left(N_{\rm max}-n\right)-1](1-e^{-\lambda(r_{\rm ph})})
\nonumber
\\
&=2N_{\rm max}e^{-\lambda(r_{\rm ph})}-(d-2N_{\rm max}-1)(1-e^{-\lambda(r_{\rm ph})})-2ne^{-\lambda(r_{\rm ph})}-2n(1-e^{-\lambda(r_{\rm ph})})
\nonumber
\\
&=2N_{\rm max}e^{-\lambda(r_{\rm ph})}-(d-2N_{\rm max}-1)(1-e^{-\lambda(r_{\rm ph})})-2n<0~.
\end{align}
Hence, if the maximum order of the Lovelock polynomial satisfies the inequality \ref{maxineq}, it is guaranteed that \ref{condition_extra} will hold identically. Thus we obtain the following condition on the metric function $e^{-\lambda}$ on the photon sphere,
\begin{align}\label{finalineq}
e^{-\lambda(r_{\rm ph})} \leq \frac{(d-2N_{\rm max}-1)}{(d-1)}~;
\qquad 1-e^{-\lambda(r_{\rm ph})} \geq \frac{2N_{\rm max}}{(d-1)}~.
\end{align}
The next step involves writing down the metric coefficient $e^{-\lambda}$ in terms of the mass of the black hole and matter outside the horizon. However, unlike the previous scenarios, in the present context the solution for $e^{-\lambda}$ can not be obtained from \ref{genL_eq_01} in a closed form, rather it yields an algebraic equation satisfied by $(1-e^{-\lambda})$, which reads,
\begin{align}\label{sol_mcomp}
\sum _{m}\hat{\alpha}_{m}r^{d-2m-1}(1-e^{-\lambda})^{m}=2m(r)~;
\qquad
m(r)\equiv M_{\rm H}+4\pi \int^{r}_{r_{\rm H}} r'^{(d-2)}\rho(r')dr'~.
\end{align}
This is the best one can do in the context of general Lovelock theories. To proceed further, one needs to solve for $e^{-\lambda(r_{\rm ph})}$ from \ref{sol_mcomp} and substitute the same in \ref{finalineq} in order to obtain the bound on the radius of the photon circular orbit. Thus for general Lovelock theories it is difficult to obtain a closed form expression for the bound on the radius of the photon circular orbit. However, if \ref{sol_mcomp} can be solved in certain special cases, e.g., in the case of Einstein-Gauss-Bonnet gravity, bound on the photon circular orbit may be obtained.

\section{Implications for quasi-normal modes and black hole shadow}\label{Ph_App}

So far we have outlined the key steps in deriving a bound (not necessarily an upper bound) on the radius of the photon sphere and have determined the same in general relativity and beyond. In particular, we have shown that in the context of braneworld black holes we have a lower bound on the photon circular orbit, rather than an upper bound, as general relativity predicts. It is interesting to ask for possible implications of these bounds for black hole quasi-normal modes and the black hole shadow, since both of them are intimately connected with the location of the photon sphere. 

First of all, in the eikonal approximation (i.e., in the large angular momentum limit), the quasi-normal modes of a static and spherically symmetric spacetime are given by \cite{Cardoso:2008bp,Li:2021zct},
\begin{align}\label{qnm}
\omega_{\rm QNM}=\ell \Omega_{\rm ph}-i\left(n+\frac{1}{2}\right)|\lambda_{\rm ph}|~;\qquad
\Omega_{\rm ph}=\sqrt{\frac{e^{\nu(r_{\rm ph})}}{r_{\rm ph}^{2}}}~,\qquad
\lambda_{\rm ph}=\sqrt{-\frac{r_{\rm ph}^{2}}{2e^{\nu(r_{\rm ph})}}\left(\dfrac{d^{2}}{dr_{*}^{2}}\frac{e^{\nu}}{r^{2}}\right)_{r_{\rm ph}}}~,
\end{align}
where, $\Omega_{\rm ph}$ is the angular velocity at the photon sphere located at $r=r_{\rm ph}$ and $\lambda_{\rm ph}$ is the Lyapunov exponent on the photon sphere, with $r_{*}$ being the tortoise coordinate. Further, $\ell$ is the angular momentum and $n$ is the order of the quasi-normal modes. Since, we have derived bounds on the metric components and the location of the photon sphere, it follows that similar bound may appear on the quasi-normal mode frequencies as well. However, $\lambda_{\rm ph}$ depends on the derivatives of the metric coefficients and hence no such bound can be derived for the same. 

It is also possible to provide a bound on the shadow radius arising out of the location of the photon circular orbit for these static and spherically symmetric compact objects. Due to spherical symmetry, it follows that the shadow will be circular, with the shadow diameter being given by, 
\begin{align}
D_{\rm shadow}=\frac{2}{\Omega_{\rm ph}}~.
\end{align}
Thus we observe that the real part of the quasi-normal mode frequency is related to the shadow radius under eikonal approximation, by a simple algebraic relation,
\begin{align}
D_{\rm shadow}\left(\textrm{Re}~\omega_{\rm QNM}\right)=2\ell~.
\end{align}
This suggests that larger the real part of the quasi-normal mode frequencies are, smaller is the shadow radius and vice versa. It is intriguing that two apparently disjoint physical characteristics associated with the compact objects, namely the quasi-normal modes arising from the perturbation of the compact objects and the shadow radius, associated with scattering cross-section of the compact object, are indeed related with one another. This in turn suggests that possible bound on the angular velocity of a photon on the photon circular orbit will translate to respective bounds for both the real part of the quasi-normal modes as well as the shadow radius. It is worthwhile to mention that these bounds on the real part of the quasi-normal mode frequencies and the shadow radius requires both the weak energy condition, as well as the negative trace condition to be identically satisfied. Since the bound on the photon circular orbit was derived using these energy conditions in the first place. 

\paragraph{Bound for pure Lovelock theories} For generality we will derive the respective bound for pure Lovelock theories, since one can apply the results to any order of the Lovelock Lagrangian and in any number of spacetime dimensions. We know from \ref{qnm} that, 
\begin{align}
\Omega_{\rm ph}&=\sqrt{\frac{e^{\nu(r_{\rm ph})}}{r_{\rm ph}^{2}}}=\sqrt{\frac{e^{\nu(r_{\rm ph})+\lambda(r_{\rm ph})}}{r_{\rm ph}^{2}}}e^{-\lambda(r_{\rm ph})/2}
\nonumber
\\
&<\sqrt{\frac{e^{\nu(r_{\rm ph})+\lambda(r_{\rm ph})}}{r_{\rm ph}^{2}}}\sqrt{\frac{d-2N-1}{d-1}}<\frac{1}{r_{\rm H}}\sqrt{\frac{d-2N-1}{d-1}}~,
\end{align}
where in the last line we have used the result, $r_{\rm ph}>r_{\rm H}$ and the fact that $e^{\nu(r_{\rm ph})+\lambda(r_{\rm ph})}\leq 1$. Thus for general relativity in four spacetime dimensions, we obtain, $\Omega_{\rm ph}r_{\rm H}<(1/\sqrt{3})$. Similarly, for $N$th order pure Lovelock gravity in $d=3N+1$ dimensions, we obtain the bound on $\Omega_{\rm ph}$ to be identical to the one for four dimensional general relativity. Thus the bound on $\Omega_{\rm ph}$ can be translated to a corresponding bound for $\textrm{Re}~\omega_{\rm QNM}$, which reads,
\begin{align}
\textrm{Re}~\omega_{\rm QNM}=\ell \Omega_{\rm ph}<\frac{\ell}{r_{\rm H}}\sqrt{\frac{d-2N-1}{d-1}}~.
\end{align}
On the other hand, the corresponding bound on the angular diameter of the shadow takes the following form,
\begin{align}
\theta_{\rm shadow}=\frac{D_{\rm shadow}}{D_{\rm obs}}=\frac{2\ell}{D_{\rm obs}}\frac{1}{\textrm{Re}~\omega_{\rm QNM}}>\frac{2r_{\rm H}}{D_{\rm obs}}\sqrt{\frac{d-1}{d-2N-1}}~.
\end{align}
where $D_{\rm obs}$ gives the distance between the shadow and the observer. For four dimensional general relativity, the above bounds translate into, $\textrm{Re}~\omega_{\rm QNM}<(\ell/\sqrt{3}r_{\rm H})$ and $\theta_{\rm shadow}>(2\sqrt{3}r_{\rm H}/D_{\rm obs})$. For $N$th order Lovelock polynomial in $d=3N+1$ dimensions, we obtain the bounds on the real part of the quasi-normal mode frequency and shadow radius to be identical to that of four dimensional general relativity, illustrating the indistinguishability of these scenarios through physical characteristics of compact objects. Thus for any accreting matter source satisfying weak energy condition the angular diameter of the shadow will be larger than that predicted by general relativity. 

\paragraph{Bound in the braneworld scenario} In the braneworld scenario, on the other hand, the bound on the photon circular orbit is the other way around, i.e., we have $r_{\rm ph}>3M_{\rm H}$. In this case, the angular velocity on the photon circular orbit becomes bounded from below, such that, $\Omega_{\rm ph}r_{\rm ph}>(1/\sqrt{3})$. Hence the corresponding bound on the real part of the quasi-normal mode frequency and the angular diameter of the shadow becomes, 
\begin{align}
\textrm{Re}~\omega_{\rm QNM}>\frac{\ell}{\sqrt{3}r_{\rm ph}}~;\qquad
\theta_{\rm shadow}<\frac{2\sqrt{3}r_{\rm ph}}{D_{\rm obs}}~.
\end{align}
Thus the bounds on real part of the quasi-normal modes and the angular diameter of the shadow are opposite to those of pure Lovelock theories. In particular, presence of accreting matter demands larger quasi-normal mode frequencies and smaller shadow radius. 

\paragraph{Bound in Lovelock theories of gravity} Finally, for general lovelock theories of gravity, even though a bound on the radius of the photon circular orbit cannot be derived, it is possible to arrive at a bound on the real part of the quasi-normal mode frequencies and the shadow radius, thanks to the bound on $e^{-\lambda(r_{\rm ph})}$, presented in \ref{finalineq}. Thus the bounds becomes,
\begin{align}
\textrm{Re}~\omega_{\rm QNM}<\frac{\ell}{r_{\rm H}}\sqrt{\frac{d-2N_{\rm max}-1}{d-1}}~;\qquad
\theta_{\rm shadow}>\frac{2r_{\rm H}}{D_{\rm obs}}\sqrt{\frac{d-1}{d-2N_{\rm max}-1}}~.
\end{align}
Here, $N_{\rm max}$ corresponds to the maximum non-trivial order of the Lovelock polynomial allowed in a gravitational theory, living in $d$ spacetime dimensions. For Einstein-Gauss-Bonnet theory, one may improve these bounds quite a bit by using the bound on $e^{-\lambda(r_{\rm ph})}$, presented in \ref{boundegb}. Thus in this case also with accreting matter, the quasi-normal mode frequencies decrease, while the shadow radius increases. 

\section{Concluding remarks}\label{Ph_conc}

Photon sphere plays a central role in the determination of both the quasi-normal mode and the angular diameter of the shadow of compact objects. Additionally, irrespective of the nature of the central compact object, i.e., whether it is a black hole or an ECO, the photon sphere always exists and hence is a universal property of any ultra-compact objects in general relativity and beyond. Following which, we have studied the bound on the location of the photon circular orbit in higher-dimensional general relativity, braneworld gravity and in pure and general Lovelock theories, with accreting matter satisfying certain energy conditions. Except for the case of braneworld gravity, where, the presence of higher dimensions induce certain energy momentum tensor on the brane violating the energy conditions, we obtain certain upper bound on the location of the photon circular orbit, depending on the spacetime dimensions and the order of the Lovelock polynomial. For the braneworld gravity, on the other hand, in the absence of external matter fields, one arrives at a lower bound on the location of the photon circular orbit. Thus there is a crucial distinction between the higher dimensional general relativity and Lovelock theories with the braneworld models, as far as the location of the photon circular orbit is concerned. The braneworld model predicts a larger value of the radius of the photon sphere compared to the other scenarios. Also note that the scalar-tensor theories of gravity are already included in this analysis, since it can be modelled as Einstein gravity with a matter field having energy momentum tensor $T^{\mu}_{\nu}=\textrm{diag.}(-\rho,p,p_{\rm T},p_{\rm T},\cdots)$. This scenario yields an upper bound on the location of the photon circular orbit, such that in four-dimensions $r_{\rm ph}\leq 3\mathcal{M}$, where $\mathcal{M}$ is the ADM mass of the spacetime, which involves contribution from the energy density of the scalar field as well. 

These results have consequences in two of the most important observables --- (a) the real part of the quasi-normal mode frequencies and (b) the angular diameter of the shadow. From our analysis it follows that, except for the braneworld scenario, the real part of the quasi-normal mode frequencies are bounded from above, while the angular diameter of the shadow are bounded from below. Thus in the presence of accreting or, any additional matter field, the shadow diameter will increase and the real part of the quasi-normal mode frequency will decrease, which is an observable effect. On the other hand, for braneworld scenario, presence of accreting matter or, any other matter field in the spacetime will decrease the angular diameter of the photon sphere and will increase the real part of the quasi-normal mode frequencies. This is expected, since the violation of the energy conditions by the extra-dimensional contribution is tamed by additional matter fields respecting energy conditions. 

The results presented here also opens up several future directions, which needs to be explored. First of all, the bounds on the photon circular orbits, derived here, are for static and spherically symmetric spacetimes. It would be interesting to ask for possible generalizations to rotating spacetimes as well. The procedure, as outlined here, can first be applied to slowly rotating spacetime, using the Hartle-Thorne approximation, before generalizing to arbitrarily rotating spacetime. Secondly, the above derivation of the bound on the location of the photon circular orbit parallels very much the derivation of the Buchdahl limit. It would be interesting to derive possible connection between the two. Additionally, if these bounds can be realized and possibly verified using astrophysical observations, one can comment on the nature of the matter distribution falling within the black hole. Similarly, the analysis presented here is based on spherical topology for the transverse sector and thus it would be interesting to ask what happens to the bound on the photon circular orbit if the topology of the transverse sector is different. It would also be important to ask whether the bound derived in this work can be coupled with the Penrose's inequality in order to arrive at certain bounds for the size of the black hole as well. This was attempted within the purview of general relativity in \cite{Yang:2019zcn,Lu:2019zxb} (also see \cite{Ma:2019ybz}) and hence it would be interesting to find out possible generalization to alternative theories of gravity in the spirit of the present work. Also in \cite{Yang:2019zcn}, the energy conditions necessary to derive this bound on the photon circular orbit was relaxed. Instead of both the weak energy condition and the negative trace condition, only the null energy condition was found sufficient to derive the bound on the photon circular orbit in general relativity. Following which, possible relaxation of the energy conditions for alternative theories of gravity must also be seek for. Finally, the presence of negative pressure at the horizon, can also be generalized to the cosmological horizon, which may describe the origin of the late time cosmic acceleration. We leave these issues for future. 
\section*{Acknowledgements}

Research of S.C. is funded by the INSPIRE Faculty fellowship from DST, Government of India (Reg. No. DST/INSPIRE/04/2018/000893) and by the Start-Up Research Grant from SERB, DST, Government of India (Reg. No. SRG/2020/000409). 
\bibliography{References}

\bibliographystyle{./utphys1}
\end{document}